# Robust intensification of global ocean Eddy Kinetic Energy from three decades of satellite altimetry observations

Bàrbara Barceló-Llull[1][*], Pere Rosselló[1], Vincent Combes[1], Antonio Sánchez-Román[1], M. Isabelle Pujol[2] and Ananda Pascual[1]

[1]Institut Mediterrani d'Estudis Avançats, IMEDEA (CSIC-UIB); Esporles, Spain.

[2]Collecte Localisation Satellites; Ramonville-Saint-Agne, France.

*Corresponding author. Email: bbarcelo@imedea.uib-csic.es

**Abstract:**

Ocean mesoscale variability, a key component of the climate system, influences ocean circulation and heat, gas, carbon and nutrient distribution. Trends on Eddy Kinetic Energy (EKE), a metric measuring its intensity, are investigated using two products constructed from 30 years of altimetric observations. Statistically significant positive trends in globally-averaged EKE reveal a strengthening of 1-3% per decade. Regions of intense mesoscale activity become more energetic than other areas. Robust positive EKE trends are observed in the Kuroshio Extension and the Gulf Stream, with remarkable EKE increases of ~50% and ~20%, respectively, over the last decade. Our study opens a new question into how the observed Gulf Stream strengthening impacts the AMOC, and challenges existing climate models emphasizing the necessity for improved small-scale ocean process representation.

## Main Text:

### Introduction

Earth has undergone anthropogenic global warming driven by the emissions of greenhouse gasses that have been released to the atmosphere by human activities since the beginning of the Industrial Revolution (*1*). The ocean is the major Earth's heat reservoir and has absorbed 90% of the anthropogenic excess heat (*2*). Ocean circulation plays a key role in the global climate system by redistributing water masses and their properties, including heat and carbon, throughout the global ocean. At the same time, changes in the climate system have not only warmed the upper ocean but also altered the wind stress, heat, and freshwater fluxes that act as driving forces for ocean circulation ([2], [3]). Hence, climate change can modify the intricate system that constitutes the global ocean circulation ([4-6]). The mesoscale circulation, defined as those motions with spatial scales between ~10-100 km, is an essential component of the global ocean circulation and is in many ways dynamically analogous to atmospheric weather ([7]). It is constituted by a time-mean (or steady) flow and a time-varying flow, which we will refer here as the mesoscale variability and includes fronts, meanders and eddies. These features exist throughout the global ocean and transport and mix water masses and their properties over long distances and locally in depth, influencing larger and smaller-scale processes ([7], [8]). The kinetic energy associated with the mesoscale variability (called Eddy Kinetic Energy, EKE) represents about 90% of the total kinetic energy of the oceans ([9], [10]), making these features an essential component of the ocean circulation. Here, we aim to evaluate if the global ocean mesoscale variability is changing over the altimetric era through the analysis of global trends in EKE, which is a metric commonly used to determine the intensity of ocean currents. Providing knowledge on this topic is crucial for predicting the future state of our oceans and their role in the broader climate system.

Recently, several investigations have been exploring this subject using datasets that may have limitations for long-term analyses ([11]-[14]). These studies employ data products with an inconsistent number of observations over time, potentially introducing erroneous trends to the evaluated variables ([15], [16]). The ocean currents of these products are derived from sea level measurements collected over the last three decades by a constellation of satellite altimeters with a varying number of missions that range from two to seven (Extended Data Fig. 1). The inclusion of new satellites over time enhances the capacity to map mesoscale structures, despite the time-variable errors dependent on the number of satellites used. The altimetric gridded product obtained from these observations has been essential to understand the large-scale and mesoscale ocean circulation ([17], [18]), and a continuous effort is done to improve its resolution and accuracy, for instance with updated corrections ([19], [20]) and mapping parameters or techniques ([21]-[23]; among others), and with new satellites such as the recently launched Surface Water and Ocean Topography (SWOT) mission ([24], [25]). An additional altimetric product is constructed for climate applications ([22]). This product includes observations from a consistent pair of altimeters, which is considered the minimum requirement for resolving mesoscale features ([16]), and prioritizes the stability of the global mean sea level assuming the cost of reducing the spatial resolution. A constant number of altimeters ensures almost uniform errors over the altimetric era, with minor variations due to changes in the satellite constellation (*14*). In consequence, this altimetric product is focused on the analysis of the long-term evolution of sea level, being appropriate for climate applications (*22*).

This study aims to assess if the global ocean mesoscale variability is becoming more energetic through the analysis of 30 years of altimetric data (1993-2022), paying special attention to regions characterized by intense mesoscale activity (Fig. 1). To address this objective, we evaluate global trends of EKE computed from altimetric observations provided by the climatic product (herein *two-sat*) and also by the product that includes all available altimeter missions (herein *all-sat*) to estimate the impact of the increasing number of satellites on the results.

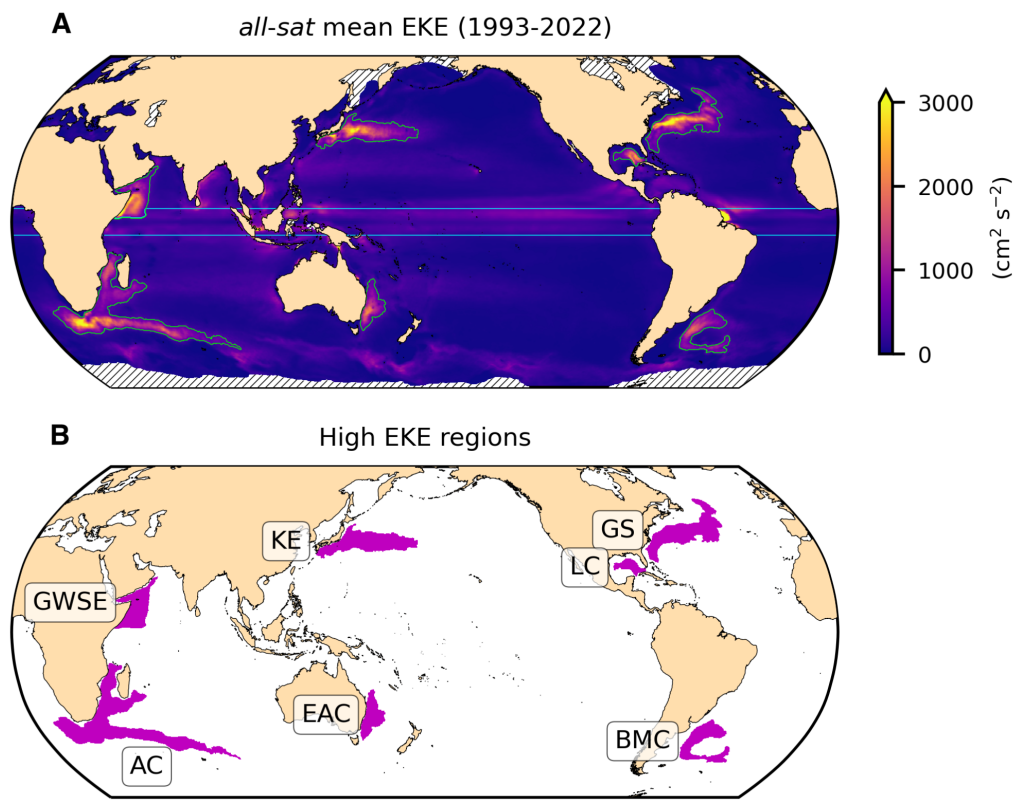

**Fig. 1. Mean EKE and high EKE regions.** (**A**) Global map of the temporally averaged EKE field computed over the period 1993-2022 at each grid point (1/4º of spatial resolution) for the *all-sat* altimetric product. Green contours outline the high EKE regions

and cyan lines delimit the tropical band. Black oblique lines indicate the areas masked due to ice coverage throughout the annual cycle. (**B**) Map showing the high EKE regions. GWSE: Great Whirl and Socotra Eddy in East Africa, AC: Agulhas Current, KE: Kuroshio Extension, EAC: East Australian Current, GS: Gulf Stream, LC: Loop Current, BMC: Brazil-Malvinas Confluence region.

**Global Eddy Kinetic Energy trends**

The trends of the globally-averaged EKE time series over the altimetric era (1993-2022) are positive and statistically significant at the 95% confidence level for both altimetric products, with a large difference in magnitude (Fig. 2A, C). The trend of the *all-sat* EKE time series is 0.64 cm$^2$ s$^{-2}$ y$^{-1}$ (or 0.066 J m$^{-3}$ year$^{-1}$), while the trend of the *two-sat* EKE time series is 0.19 cm$^2$ s$^{-2}$ y$^{-1}$ (or 0.019 J m$^{-3}$ year$^{-1}$), i.e., 3.4 times smaller (Fig. 3C). These EKE trends represent an increase per decade of 2.8% for *all-sat* and 0.8% for *two-sat*, relative to their respective mean EKE values (Extended Data Table 1). Assuming an area of the surface global ocean of 3.28 × 10$^8$ km$^2$, the area-integrated EKE trend is 0.22 × 10$^{15}$ J m$^{-1}$ decade$^{-1}$ for the *all-sat* product and 0.06 × 10$^{15}$ J m$^{-1}$ decade$^{-1}$ for the *two-sat* product. The result obtained from the *all-sat* altimetric product may suggest that the ocean mesoscale variability is experiencing a strong intensification (*11*, *13*). However, the different result obtained from the *two-sat* product, which is based on the observations gathered by a consistent number of satellites, suggests that the larger *all-sat* EKE trend may be, at least partially, an artifact induced by the increasing number of satellites in the altimetric record (*12*, *14*) (Extended Data Fig. 1). The inclusion of additional satellites within the altimetric record can enhance the capacity to detect higher energy levels. Consequently, a progressive increase in the satellite count over time may give rise to an increase of energy attributed to this phenomenon. Another reason for this difference could be that *two-sat* may not completely capture a potential increase in mesoscale kinetic energy due to its lower resolution. Hence, both altimetric products support the conclusion that the global ocean mesoscale variability is becoming more energetic over the altimetric era, although the magnitude may be overestimated by *all-sat* and underestimated by *two-sat*.

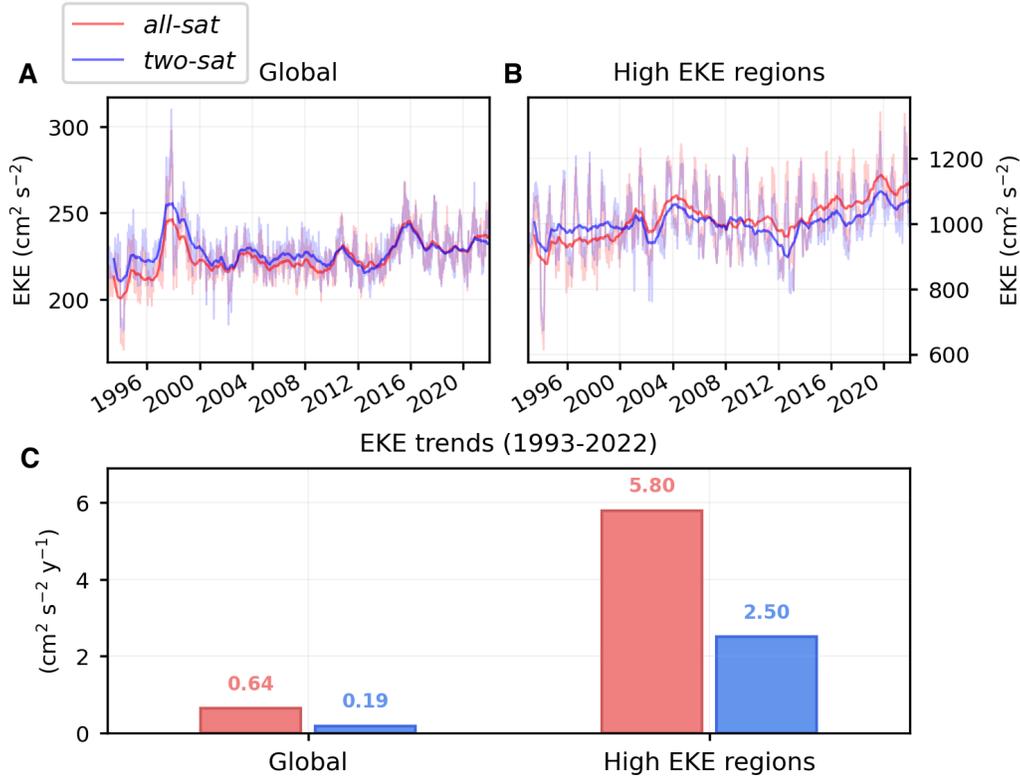

**Fig. 2. EKE time series and trends.** Area-weighted mean EKE time series computed over (**A**) the global ocean and (**B**) the high EKE regions, for the *all-sat* (red line) and *two-sat* (blue line) altimetric products. Thinner lines represent the original data, while thicker lines show the yearly-rolling mean (i.e. 365-day-window moving average). (**C**) Trends of the original area-weighted mean EKE time series shown in (A, B), computed from 1993 to 2022. All trends are statistically significant ($p < 0.05$).

A sensitivity test reveals that the ratio between the global *all-sat* EKE trend and the global *two-sat* EKE trend is reduced from 3.4 between 1993-2022 to 1.8 when computed for the period 2000-2022, and this value is maintained or slightly smaller for the periods evaluated afterwards (Fig. 3A, C). The year 2000 marks the division between two distinct periods in the altimetry era: an initial period characterized by a varying number of satellites ranging from 2 to ~3, followed by a second period where the number of satellites was consistently higher than 2 (with the exception of one month in 2008) (Extended Data Fig. 1). This result indicates that the difference between *all-sat* and *two-sat* EKE trends is partly related to the varying number of satellites included to build the *all-sat* product, and this difference is higher when computing trends over the complete altimetry era and reduced, but still important, when computing trends after 2000.

A striking feature is that *two-sat*, whose limitations are not time-dependent, show a clear increase of the global EKE trend for the periods evaluated in the sensitivity test (Fig. 3A). Considering the complete altimetric era, the global *two-sat* EKE trend is 0.19 cm$^2$ s$^{-2}$ y$^{-1}$, but this trend gets larger when shortening the time series, to a maximum of 1.19 cm$^2$ s$^{-2}$ y$^{-1}$ over 2012-2022. Hence, both altimetric products reveal an intensification of the global ocean mesoscale variability, and *two-sat* demonstrates that this increase in EKE is much faster when evaluating the last two decades. The EKE trend computed over 2012-2022 for *two-sat* is 6.2 times larger than the trend computed over 1993-2022, while for *all-sat* this ratio is only 2.2 (Fig. 3A). This difference is related to the impact that the number of satellites included in the *all-sat* product has on the computation of EKE trends. When considering the complete altimetric era, the number of satellites drastically changes from ~2 satellites between 1993-2000, to ~3-4 satellites between 2000-2016, and to higher than 4 satellites after 2016 (Extended Data Fig. 1). This increase over time on the number of satellites implies an enhancement on the capacity to detect energy. A progressive increase on the satellite count could lead to an increase of the energy observed related to this phenomenon, rather than a real increase of energy in the ocean. This circumstance would result in an overestimation of the EKE trends computed from *all-sat*. However, this overestimation diminishes when analyzing shorter, more recent time periods, because this phenomenon decreases if we exclude the first years of data computed from only ~2 satellites.

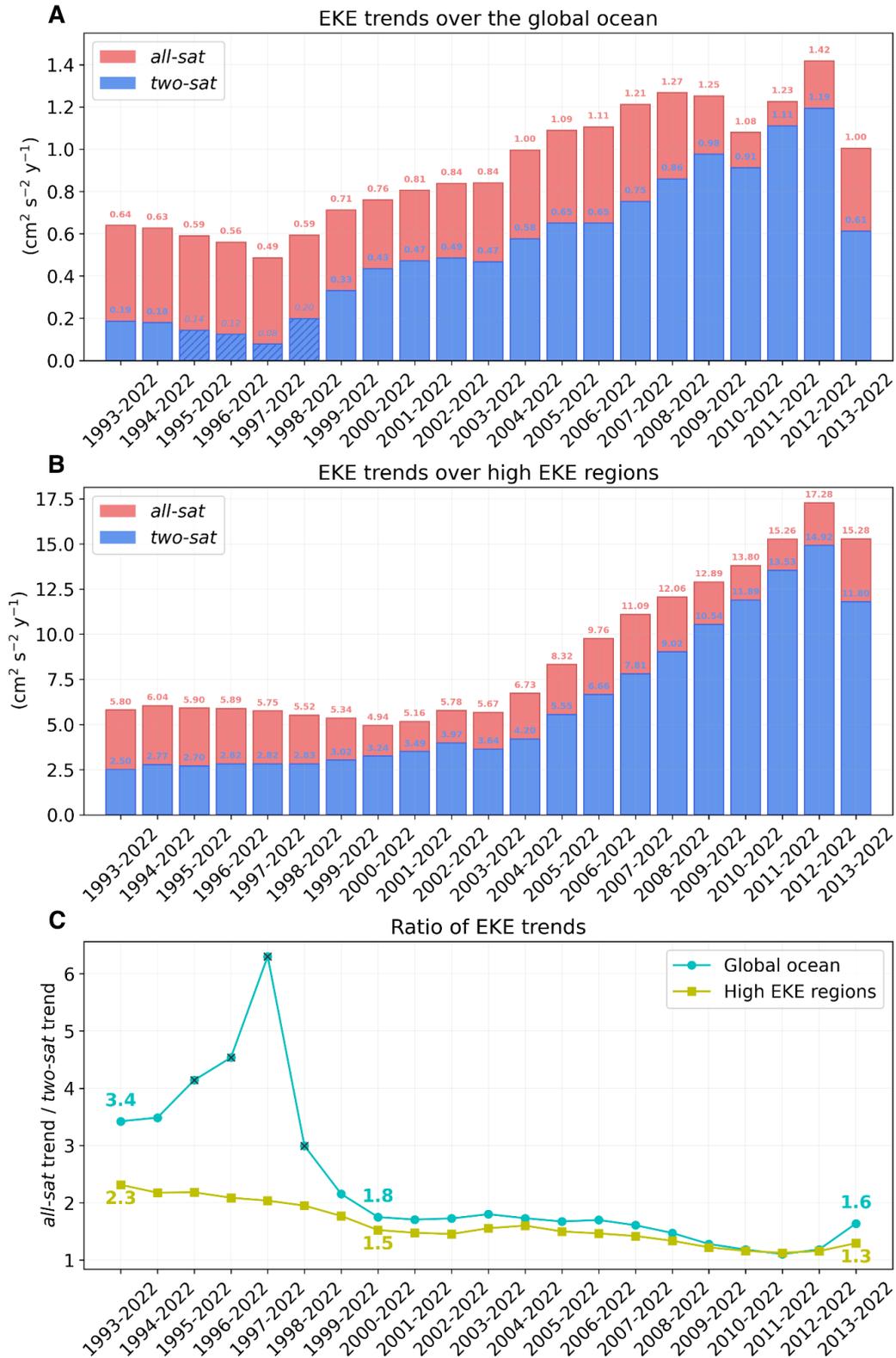

**Fig. 3. Sensitivity test.** EKE trends computed over the (**A**) global ocean and (**B**) high EKE regions for different periods. Significant trends (p < 0.05) are written in bold text,

while non-significant trends are written in italics. (**C**) Ratio of the *all-sat* EKE trend divided by the *two-sat* EKE trend.

Martínez-Moreno et al. (*13*) reported a trend of the global surface area-integrated EKE of $(0.09 \pm 0.04) \cdot 10^{15}$ J m$^{-1}$ decade$^{-1}$ using a previous version (vDT2018) of the *all-sat* altimetric product and following a similar methodology but computing trends over the period between 1 January 1993 to 7 March 2020 and from a smoothed 365-day running average time series. The *all-sat* EKE trend computed here over 1993-2022 from the currently available vDT2021 version is 2.4 times that value (Fig. 2). We have also calculated the *all-sat* EKE trend over the same period as Martínez-Moreno et al. and over a 365-day running average (Extended Data Fig. 2), without finding large differences with respect to the trends calculated from the original data over 1993-2022 (Fig. 2). This suggests that the discrepancy between the trends computed in this study and the trends reported by Martínez-Moreno et al. is related to the version of the product used. We cannot test this sensitivity because the *all-sat* vDT2018 altimetric product is no longer available. However, the *two-sat* vDT2018 altimetric product is still available on the Copernicus Climate Change Service (C3S) website ([10.24381/cds.4c328c78](10.24381/cds.4c328c78)) and we have found high sensitivity in the results related to the version of the product analyzed (Extended Data Fig. 3). Indeed, a global non-significant EKE trend of -0.002 cm$^2$ s$^{-2}$ y$^{-1}$ was obtained for the *two-sat* vDT2018 altimetric product in comparison to the 0.114 cm$^2$ s$^{-2}$ y$^{-1}$ non-significant trend obtained for the same period with *two-sat* vDT2021 (Extended Data Fig. 3).

**Eddy Kinetic Energy trends over regions of intense mesoscale activity**

The mesoscale variability intensification is clearly concentrated on regions characterized by high EKE levels (Fig. 1, 2, Extended Data Fig. 4). In contrast, the Tropics and the rest of the ocean exhibit predominantly statistically non-significant trends in the averaged EKE time series (Extended Data Fig. 4). The distinct peaks detected in global and tropical time series, approximately corresponding to 1998 and 2016, were previously identified as El Niño events (*13*). The EKE trend computed over high EKE regions from *all-sat* is 5.80 cm$^2$ s$^{-2}$ y$^{-1}$ (or 0.59 J m$^{-3}$ year$^{-1}$), while from *two-sat* is 2.50 cm$^2$ s$^{-2}$ y$^{-1}$ (or 0.26 J m$^{-3}$ year$^{-1}$) (Fig. 2B, C), i.e., 2.3 times smaller (Fig. 3C). These EKE trends represent an increase per decade of 5.7% for *all-sat* and 2.5% for *two-sat*, relative to their respective mean EKE values (Extended Data Table 1). Assuming a surface area of $1.65 \times 10^7$ km$^2$, the area-integrated EKE trend is $0.10 \times 10^{15}$ J m$^{-1}$ decade$^{-1}$ for *all-sat* and $0.04 \times 10^{15}$ J m$^{-1}$ decade$^{-1}$ for *two-sat*. A previous study reported an EKE increase rate of 2.5% per decade from *all-sat* vDT2018, with a different definition of high EKE regions ([13]).

In high EKE regions the sensitivity test of *two-sat* shows a progressive increase of the EKE trends computed over the periods evaluated (Fig. 3B, blue bars). The sensitivity test of *all-sat* also shows a progressive increase of EKE trends (Fig. 3B, red bars), but this increase is smaller due to the time-dependent limitations of *all-sat* explained in the previous section. Because of this, we focus here the discussion on the results obtained from *two-sat*. Over the last 20 years the *two-sat* EKE trend is slightly higher than that computed over the complete 30 years of altimetry data, but afterwards the trend increases to a maximum of 14.9 cm$^2$ s$^{-2}$ y$^{-1}$ for the period 2012-2022, which is 6.0 times larger than the trend computed over 1993-2022. This means that the ocean mesoscale variability is becoming more energetic in high EKE regions than in other regions of the world ocean (Extended Data Fig. 4), and the pace of this increase of energy is faster over the last decade (Fig. 3).

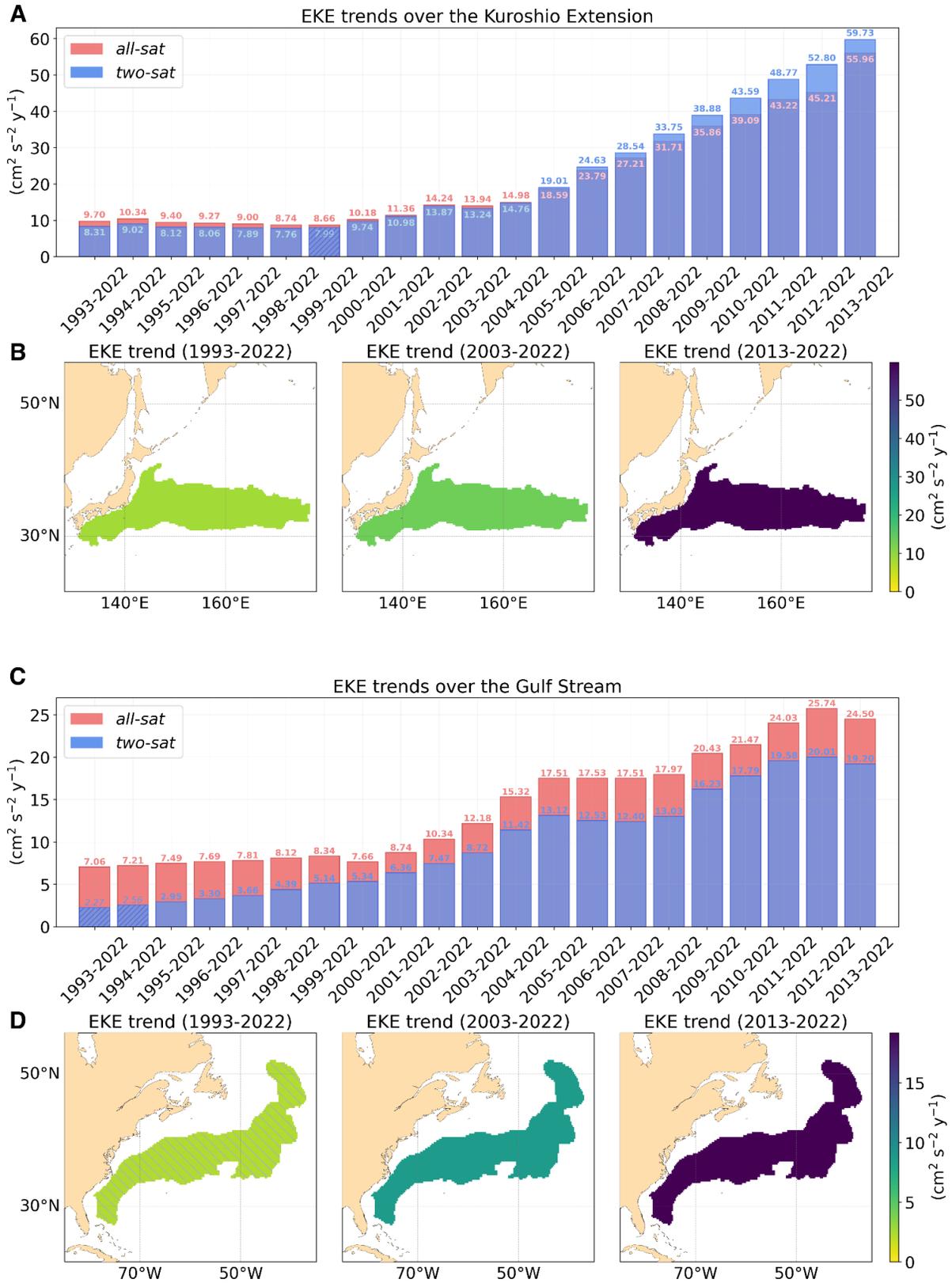

**Fig. 4. Sensitivity test over the Kuroshio Extension and the Gulf Stream.** EKE trends computed over (**A**) the Kuroshio Extension (KE) and (**C**) the Gulf Stream (GS) for different periods. Trends are computed from the original area-weighted mean EKE time series shown in Extended Data Fig. 5. Significant trends (p < 0.05) are written in bold text, while non-significant trends are written in italics. Maps showing the *two-sat* EKE trends over (**B**) the Kuroshio Extension and (**D**) the Gulf Stream for the last 30, 20 and 10 years. Oblique gray lines show non-statistically significant trends (p < 0.05).

An examination of specific high EKE regions (Extended Data Figs. 5 and 6, and Fig. 4) reveals that the domains with statistically significant positive EKE trends for both altimetric products and maintained over the periods evaluated in the sensitivity test are the Kuroshio Extension (Fig. 4A) and the Gulf Stream (Fig. 4C; with the exception of the *two-sat* 1993-2022 and 1994-2022 trends, that are not statistically significant). In the Kuroshio Extension, with both altimetric products we obtain similar EKE trends, being the ratio between the *all-sat* trend and the *two-sat* trend 1.2 for the period 1993-2022 and 0.9 between 2013 and 2022 (Fig. 4A). This suggests that in this region the impact of the varying number of satellites used to build *all-sat* is smaller than in the other regions, or even negligible, and that *two-sat*, with smaller resolution, is able to capture the magnitude of the increasing trends. Over the last 30 years, the EKE trends in the Kuroshio Extension are 9.70 cm$^2$ s$^{-2}$ y$^{-1}$ for *all-sat* and 8.31 cm$^2$ s$^{-2}$ y$^{-1}$ for *two-sat*, while over the last 10 years these trends are 55.96 cm$^2$ s$^{-2}$ y$^{-1}$ and 59.73 cm$^2$ s$^{-2}$ y$^{-1}$, respectively (Fig. 4A, B). This result indicates that over the last decade the EKE in the Kuroshio Extension has increased 6 to 7 times faster than over the last three decades. The EKE trends computed over the altimetric era represent an increase per decade relative to their respective mean values (Extended Data Table 1) of 9.3% for *all-sat* and 8.1% for *two-sat*, while the EKE trends computed for the last decade represent an increase of 54% and 58%, respectively. In addition, the Kuroshio Extension is the high EKE region with the largest trends from both altimetric products (Extended Data Figs. 5 and 6, and Fig. 4). A poleward migration and intensification of the Kuroshio Extension system has been previously reported from datasets that use the *all-sat* altimetric product and from climate models, revealing a relation between the EKE evolution and the Pacific Decadal Oscillation (*26*, *27*). The analysis of EKE trends conducted here corroborates the intensification detected previously in the Kuroshio Extension.

The Gulf Stream is the second region with the largest EKE trends (Extended Data Figs. 5 and 6, and Fig. 4). Over the last three decades, the EKE trends are 7.06 cm$^2$ s$^{-2}$ y$^{-1}$ for *all-sat* and a statistically non-significant 2.27 cm$^2$ s$^{-2}$ y$^{-1}$ for *two-sat*, representing a ratio of 3.1 (Fig. 4C). These trends increase to 24.50 cm$^2$ s$^{-2}$ y$^{-1}$ for *all-sat* and 19.20 cm$^2$ s$^{-2}$ y$^{-1}$ for *two-sat* (both trends are statistically significant) over the last decade, showing a reduction in the ratio to 1.3 (similar to the result obtained for the global ocean, Fig. 3C). This indicates that in the Gulf Stream the *two-sat* EKE trend has increased 8.5 times faster over the last decade than over the complete altimetric era (Fig. 3D), and that the *all-sat* EKE trend evolution may be affected by the varying number of satellites (see the discussion for the global ocean). The EKE trends computed over the last 30 years represent an increase per decade relative to their respective mean values (Extended Data Table 1) of 6.5% for *all-sat* and 2.1% (statistically non-significant) for *two-sat*, while the EKE trends computed over the last decade represent an increase of 23% and 18%, respectively. A positive, consistent and increasing EKE trend in the Gulf Stream is a new result that contrasts with previous climate studies. The Gulf Stream contributes to the Atlantic meridional overturning circulation (AMOC) and is also the western boundary current of the subtropical North Atlantic gyre circulation. The AMOC plays a crucial role in regulating Earth's climate and is constituted by the Gulf Stream, that transports warm and saline Atlantic water polewards, the cooling and densification of this water in the Nordic Seas, and the return of the cooled water at depth ([28]). Previous climate studies reported a weakening of the AMOC (e.g. [29], [30]). Our study raises a new question regarding how the observed strengthening of the Gulf Stream mesoscale variability influences the temporal evolution of the AMOC. In the Southern Ocean, experiments with high-resolution ocean models reveal that mesoscale eddies may mitigate the effects of a warming climate by maintaining the strength of the Antarctic Circumpolar Current ([31]) and delaying the decline of Antarctic sea ice ([32]). These processes, together with the increase of EKE detected in the Gulf Stream, occur at scales smaller than those resolved in climate models. To study the impact of small-scale ocean processes in the large-scale climate system we need to better represent them in climate models and projections, or use alternative approaches ([33]). In addition, the relationship between the Gulf Stream, the AMOC and the subtropical North Atlantic gyre is complex and makes necessary a sustained long-term observation of the ocean and the development of novel techniques to analyze all available data comprehensively ([34]). The study conducted here highlights the importance of analyzing the relation between the strengthening of the mesoscale variability in the Gulf Stream and the temporal evolution of the subtropical North Atlantic gyre and AMOC.

Chi et al. (*35*) evaluated 26 years of along-track altimetric data (1993–2018) to determine if the expected deceleration and poleward shift of the Gulf Stream by climate predictions were observable. They calculated linear trends in several metrics (latitude, transport, width and maximum downstream velocity) in stream-following coordinates and concluded that the trends were not significant. They also mentioned that the only locations with trend confidence showed that the Gulf Stream had accelerated and narrowed. Another study based on satellite observations revealed a correlation between mesoscale variability and the Gulf Stream meridional position (*36*). A more energetic mesoscale field was associated with a northward shift of the Gulf Stream position and with a positive North Atlantic Oscillation. The analysis of other observational data indicate a weakening of mass transport in transects that include the Gulf Stream and its recirculations (*34*, *37*, *38*). Future studies should evaluate the influence of the enhanced mesoscale variability in the mass transport across these sections.

The positive and increasing EKE trends obtained in the Kuroshio Extension and the Gulf Stream are in opposition to the non-significant trends reported by (*13*) from the *all-sat* vDT2018 altimetric product (their Fig. 6 in Extended Data). A possible reason for this difference may be the distinct methodologies followed to define the high EKE regions. They define them with the 99th percentile of the mean kinetic energy (computed from the time-mean velocity field), while we use the 90th spatial percentile on the mean EKE field. We have compared the boundaries obtained from each methodology and found negligible differences (Extended Data Fig. 7). Hence, the discrepancy between results comes from the older version of the altimetric product that they used and the shorter time series that was available at that moment. As the time series expands and the altimetric products are improved in the future, it will be necessary to reevaluate the analysis of the EKE trends conducted here to determine if our results remain consistent over time.

Beech et al. (*39*) analyzed the long-term evolution of EKE using a climate model with variable-resolution aimed at increasing grid precision in high EKE regions. That model has the knowledged limitations of (i) underrepresenting the EKE with respect to altimetric observations (especially in high latitudes) and (ii) representing a North Atlantic EKE distribution more zonal than observed by satellite altimetry (their Fig. 2). In that study EKE is projected to shift poleward in several high EKE regions, to increase in the Kuroshio Current and to decrease in the Gulf Stream. However, their EKE representation in the Kuroshio Current is more similar to altimetric observations than the EKE representation in the Gulf Stream, which is smaller in magnitude, particularly in the northern part, and differs in position. They show that the Gulf Stream is projected to decrease in eddy activity over the twenty-first century, in opposition to the result obtained here from 30 years of satellite altimetry. This discrepancy is likely attributable to the limitations of the climate model over the North Atlantic.

**Conclusions**

We have investigated the Eddy Kinetic Energy (EKE) temporal evolution to evaluate if the surface global ocean is becoming more energetic through the analysis of 30 years of satellite altimetry data (1993-2022). Ocean mesoscale variability is a key component of the global ocean circulation and includes fronts, meanders and eddies on spatial scales between ~10-100 km. The EKE associated with these features accounts for about 90% of the total kinetic energy of the oceans (*9*, *10*). We have computed EKE trends from two altimetric products: *all-sat* includes all available altimetry data and is constructed to study mesoscale dynamics, while *two-sat* considers a consistent number of satellites and is built for climate applications. The globally-averaged EKE time series over the altimetric era (1993-2022) show statistically significant positive trends, with the *all-sat* product indicating a larger increase compared to the *two-sat* product. Our results suggest that the increasing number of satellites in the altimetric record may partly contribute to the observed differences. Despite this, both altimetric products support the conclusion that the global ocean mesoscale variability is strengthening. The ocean is becoming more energetic in high EKE regions than in other regions of the world ocean, and the pace of this increase of energy is 6 times faster over the last decade. Robust statistically significant positive EKE trends are observed in the Kuroshio Extension and the Gulf Stream. The Kuroshio Extension has the largest EKE trends from both altimetric products. Over the last three decades, the EKE trends in the Kuroshio Extension are 9.70 $cm^2$ $s^{-2}$ $y^{-1}$ for *all-sat* and 8.31 $cm^2$ $s^{-2}$ $y^{-1}$ for *two-sat*, indicating an intensification of EKE of ~8-9% per decade with respect to mean values. The trends in this region are similar for both datasets, suggesting that the impact of the varying number of satellites used to build *all-sat* is smaller than in the other regions. Over the last decade, the EKE in the Kuroshio Extension has increased 6 to 7 times faster than over the last three decades, representing an increase of ~50% with respect to mean values. These findings support previous studies that detected an intensification of the Kuroshio Extension, potentially linked to the Pacific Decadal Oscillation. The Gulf Stream is the second region with the largest EKE trends. Over the altimetric era, the EKE trends in the Gulf Stream are 7.06 cm$^2$ s$^{-2}$ y$^{-1}$ for *all-sat* and a statistically non-significant 2.27 cm$^2$ s$^{-2}$ y$^{-1}$ for *two-sat* (note that for *two-sat* non-significant trends are obtained only for the periods 1993-2022 and 1994-2022, being significant for all the other periods evaluated in the sensitivity test), representing an increase per decade relative to their respective mean values of 6.5% for *all-sat* and 2.1% for *two-sat*. Over the last decade, this region has increased 8.5 times faster than over the complete altimetric era, representing a statistically significant increase of ~20% with respect to mean values. A positive, consistent and increasing EKE trend in the Gulf Stream opens new questions about its relationship with the Atlantic meridional overturning circulation (AMOC) and the subtropical North Atlantic gyre. Sustained long-term observation of the ocean and the development of novel techniques to analyze all available data exhaustively are necessary to study the complex relationship between the Gulf Stream, the AMOC and the subtropical North Atlantic gyre. The observed strengthening of mesoscale variability in the Gulf Stream challenges existing climate model projections. Our findings emphasize the need for improved representation of small-scale ocean processes in climate models to better understand their influence on the large-scale climate system. A comprehensive analysis of the dynamics driving changes in mesoscale variability is necessary to discern anthropogenic change from natural variability. As the altimetric record increases and future advancements enhance altimetric products, it will be necessary to reassess the analysis of EKE trends conducted here to verify the consistency of our findings over time.

**Data and methods**

**Altimetry data products.** In this study, we use the latest version of the global multi-satellite Delayed Time (DT) Data Unification and Altimeter Combination System (DUACS) (*14*, *19*, *40*), named vDT2021 and freely available through the European Copernicus Program (https://marine.copernicus.eu/). The vDT2021 product supersedes the previous vDT2018 version when comparing with independent in situ observations (*22*). The DUACS system generates two distinct types of altimetric Level-4 (L4) gridded products for the global ocean: the all-satellites (herein *all-sat*) and the two-satellites (herein *two-sat*) products. The *all-sat* product (*14*), disseminated via the Copernicus Marine Service (CMEMS) project[1], incorporates all available altimeters at a given time, ranging from 2 to 7 over the altimetric period (Extended Data Fig. 1). It emphasizes the mesoscale mapping capacity of the altimeter data and the stability of the overall dataset, despite the time-variable errors dependent on the number of satellites used (*14*). The *two-sat* product (*40*), distributed via the Copernicus Climate Change Service (C3S) project and also by CMEMS[2], is derived from a consistent pair of altimeters, which is considered the minimum requirement for retrieving mesoscale signals in delayed time conditions (*16*). The *two-sat* product is mainly based on the long-term TOPEX/POSEIDON/Jason orbit and completed by a second mission on the ERS/Envisat/AltiKa or the more recent Sentinel-3 orbit (*18*). This product prioritizes the stability of the global mean sea level, assuming the cost of reducing the spatial coverage of the ocean. The steady number of altimeters ensures nearly consistent errors throughout the entire time period, barring minor variations due to changes in the satellite constellation (*14*). The *two-sat* product is aimed at monitoring the long-term evolution of sea level, therefore it is appropriate for climate studies of sea level (large scale signals) (*22*).

The validation of altimetry products is a fundamental step in the DUACS data processing to assess and characterize the errors associated with the altimetry measurements (*41*). The quality of both *all-sat* and *two-sat* altimetric products is mainly assessed through the analysis of the sea level anomaly (SLA) field at different steps of the processing and through the evaluation of the SLA consistency along the tracks of different altimeters and between gridded and along-track products, in addition to comparisons with external in situ measurements (*14*).

---

[1] Product ID: SEALEVEL_GLO_PHY_L4_MY_008_047 doi: https://doi.org/10.48670/moi-00148
[2] Product ID: SEALEVEL_GLO_PHY_CLIMATE_L4_MY_008_057 doi: https://doi.org/10.48670/moi-00145

Both the *all-sat* and *two-sat* products provide geostrophic velocity anomalies derived from the gridded sea level anomaly (SLA) field, which is calculated with respect to a temporal mean of sea surface height over the same period (1993-2012; [14]). The geostrophic velocity anomalies provided by the altimetric products represent ocean currents at the surface and are computed through the application of the geostrophic approximation by using a 9-point stencil width methodology ([42]) for latitudes outside the ±5ºN band. In the equatorial band, they are computed through the Lagerloef methodology ([43]) with the β plane approximation. Both the *all-sat* and *two-sat* data products cover the period ranging from 1 January 1993 to 7 June 2023 (last accessed in March 2024) and have a spatio-temporal resolution of 1/4º and 1 day. To study the temporal evolution of the EKE we analyze data from complete years, i.e., from 1993 to 2022.

**Eddy Kinetic Energy computation.** The calculation of the Eddy Kinetic Energy (EKE) is performed with the following expression:

$$\text{EKE} = \frac{1}{2}\rho(u_a^2 + v_a^2),$$

where $\rho = 1025$ kg m$^{-3}$ is the constant approximated sea water density, and $u_a$ and $v_a$ are the zonal and meridional geostrophic velocity anomalies, respectively, provided by the altimetric products. The EKE SI units are J m$^{-3}$. However, we will be working instead with the EKE normalized by the density, whose units are cm$^2$ s$^{-2}$, as done by the altimetric community. The relation between the two conventions is a constant factor:

$$\text{EKE [J m}^{-3}] \sim 0.1025 \cdot \text{normalized EKE [cm}^2\text{s}^{-2}].$$

Hereinafter, the normalized EKE will be called EKE. The EKE is computed from the geostrophic velocity anomaly fields, and therefore it represents the kinetic energy associated with deviations from the mean oceanic flow.

To compute the mean EKE and the EKE trends, an ice mask is implemented to systematically exclude regions covered by ice throughout the annual cycle. Thus, the analysis is confined to latitudes between 65ºN and 65ºS, where the ocean remains mostly ice-free throughout the year, ensuring consistent, uninterrupted satellite altimetry measurements.

To calculate spatial averages of EKE, we compute the area-weighted arithmetic mean with the following equation:

$$\overline{\text{EKE}} = \frac{\sum_{i,j} \left[ \text{area}_{i,j} \cdot \text{EKE}_{i,j} \right]}{\sum_{i,j} \text{area}_{i,j}},$$

where $\text{area}_{i,j}$ is the area of each grid cell within the selected region, i represents indices along the longitude axis, and j represents indices along the latitude axis.

**Definition of high EKE regions.** In this study, we delineate areas characterized by high EKE, hereinafter referred to as high EKE regions (see Fig. 1). They are identified as those regions exceeding the 90th spatial percentile on the mean EKE field computed over the period 1993-2022 from the all-sat data product. To avoid the potential inclusion of small patches with high EKE, we adopt a filtering process consisting of selecting only large and well-defined regions (roughly above $4\text{x}10^5$ km$^2$ in area), resulting in high EKE regions only covering 5% of the global ocean. These regions coincide with the Gulf Stream (GS), the Kuroshio Extension (KE), the Agulhas Current (AC), the Brazil-Malvinas Confluence region (BMC), the Loop Current (LC), the Great Whirl and Socotra Eddy in East Africa (GWSE), and the East Australian Current (EAC) (Fig. 1).

**Computation of EKE trends.** EKE trends have been computed for the period 1993-2022 using the modified Mann-Kendall test proposed by (*44*). This approach deviates from the standard Mann-Kendall test in that it effectively addresses the autocorrelation often found in climate data, leading to more accurate trend estimates (*44*). The modified Mann-Kendall test was recently used by (*13*) to compute EKE trends globally.

## Data availability

The original altimetric data (*all-sat* and *two-sat* vDT2021 and *two-sat* vDT2018) will be publicly available on Zenodo after acceptance of the manuscript to allow reproducibility of this study, particularly after future updates to the data on the CMEMS website. The data generated in this study will be publicly available on Zenodo.

## Code availability

The codes to reproduce or extend this study will be publicly available on Zenodo.

## Acknowledgments

B. B.-L. and A. P. acknowledge fundings from the EuroSea project, funded by the European Union's Horizon 2020 research and innovation programme under grant agreement No 862626. B. B.-L. is supported by the Balearic Islands Government Vicenç Mut program, grant PD/008/2022. P. R. was supported by a JAE-Intro scholarship issued by the Spanish National Research Council (CSIC). V. C. acknowledges the support from the Ramón y Cajal Program (RYC2020- 029306-I) and from the European Social Fund/Universitat de les Illes Balears/Spanish State Research Agency (AEI—10.13039/501100011033). All authors acknowledge fundings from the Copernicus Marine Service Sea Level Thematic Assembly Center (SL-TAC) project, funded by the Copernicus Marine Service (CMEMS), the marine component of the Copernicus Program of the European Union, under contract No 21001L01-COP-TAC-SL-2100. This study was carried out within the framework of the activities of the Spanish Government through the "María de Maeztu Centre of Excellence" accreditation to IMEDEA (CSIC-UIB) (CEX2021-001198). This work is a contribution to the SL-TAC project, to the CSIC Interdisciplinary Thematic Platform (PTI) Teledetección (PTI-TELEDETECT), and to the ObsSea4Clim project (ObsSea4Clim contribution number 2). ObsSea4Clim is funded by the European Union. Views and opinions expressed are however those of the author(s) only and do not necessarily reflect those of the European Union or the European Research Executive Agency (REA). Neither the European Union nor the granting authority can be held responsible for them.

**Author contributions**

A. P., B. B.-L., P. R., V. C. and A. S. R. conceived the study. B. B.-L. and P. R. conducted the analyses. All authors contributed to the interpretation of the results. B. B.-L. wrote the first draft and all authors contributed to the revision of the manuscript.

**Competing interests**

Authors declare that they have no competing interests.

**Additional information**

Extended data

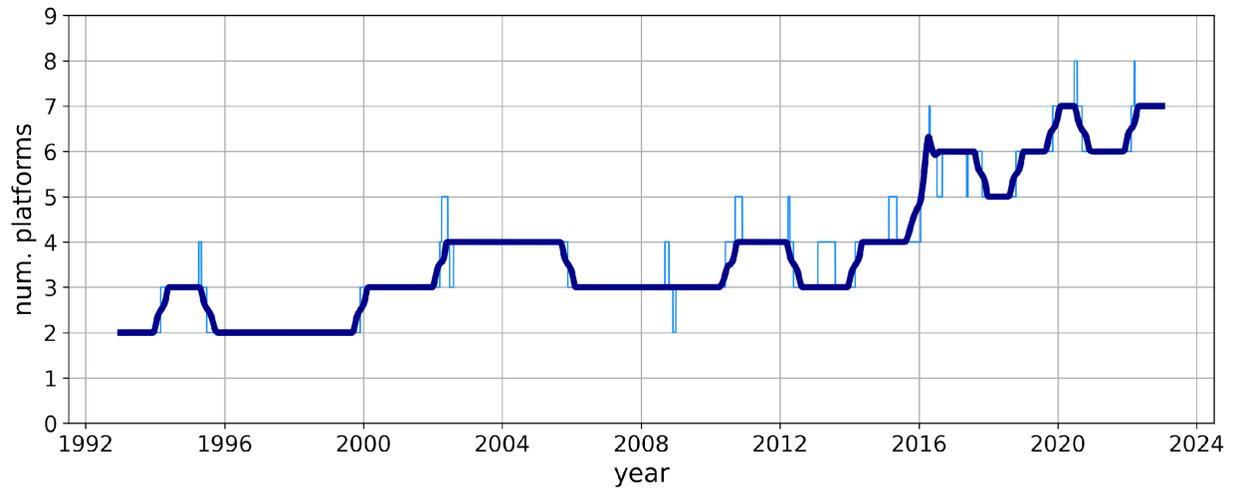

**Extended Data Fig. 1 | Number of altimetric missions.** Original (pale-blue) and yearly low-pass filtered time series (blue) of the number of satellites operating at a given time used to construct the *all-sat* altimetric product.

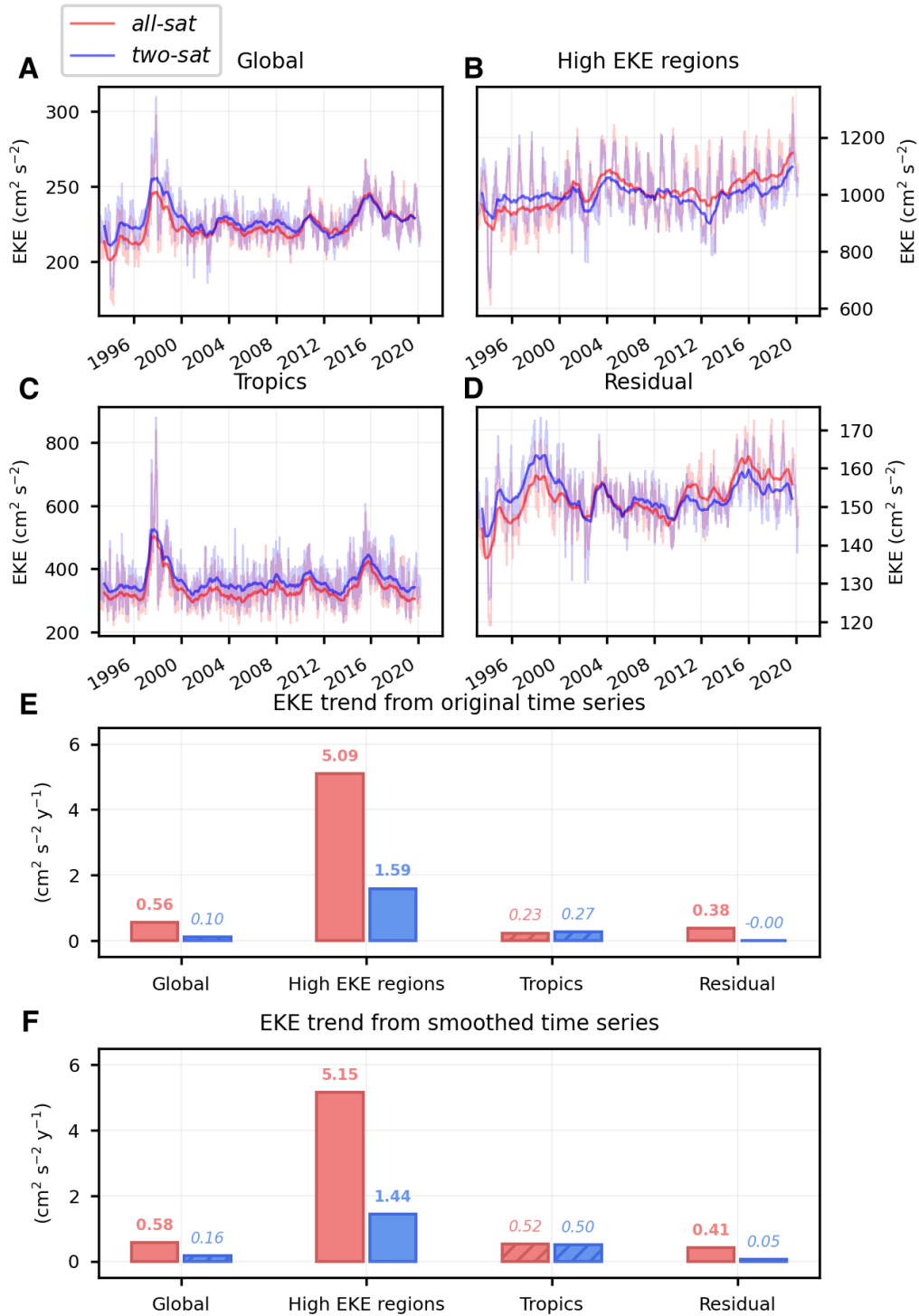

**Extended Data Fig. 2 | EKE time series and trends for the period between 1 January 1993 and 7 March 2020 (*13*).** Area-weighted mean EKE time series computed over (**A**) the global ocean, (**B**) the high EKE regions, (**C**) the tropical band, and (**D**) the global ocean excluding the high EKE regions and the tropical band (called residual), for the *all-sat* vDT2021 (red line) and *two-sat* vDT2021 (blue line) altimetric products. Thinner lines represent the original data, while thicker lines show the smoothed 365-day running average. (**E**) Trends of the original area-

weighted mean EKE time series shown in (A-D). (**F**) Trends of the smoothed area-weighted mean EKE time series shown in (A-D). In (E) and (F) significant trends ($p < 0.05$) are written in bold text, while non-significant trends are written in italics.

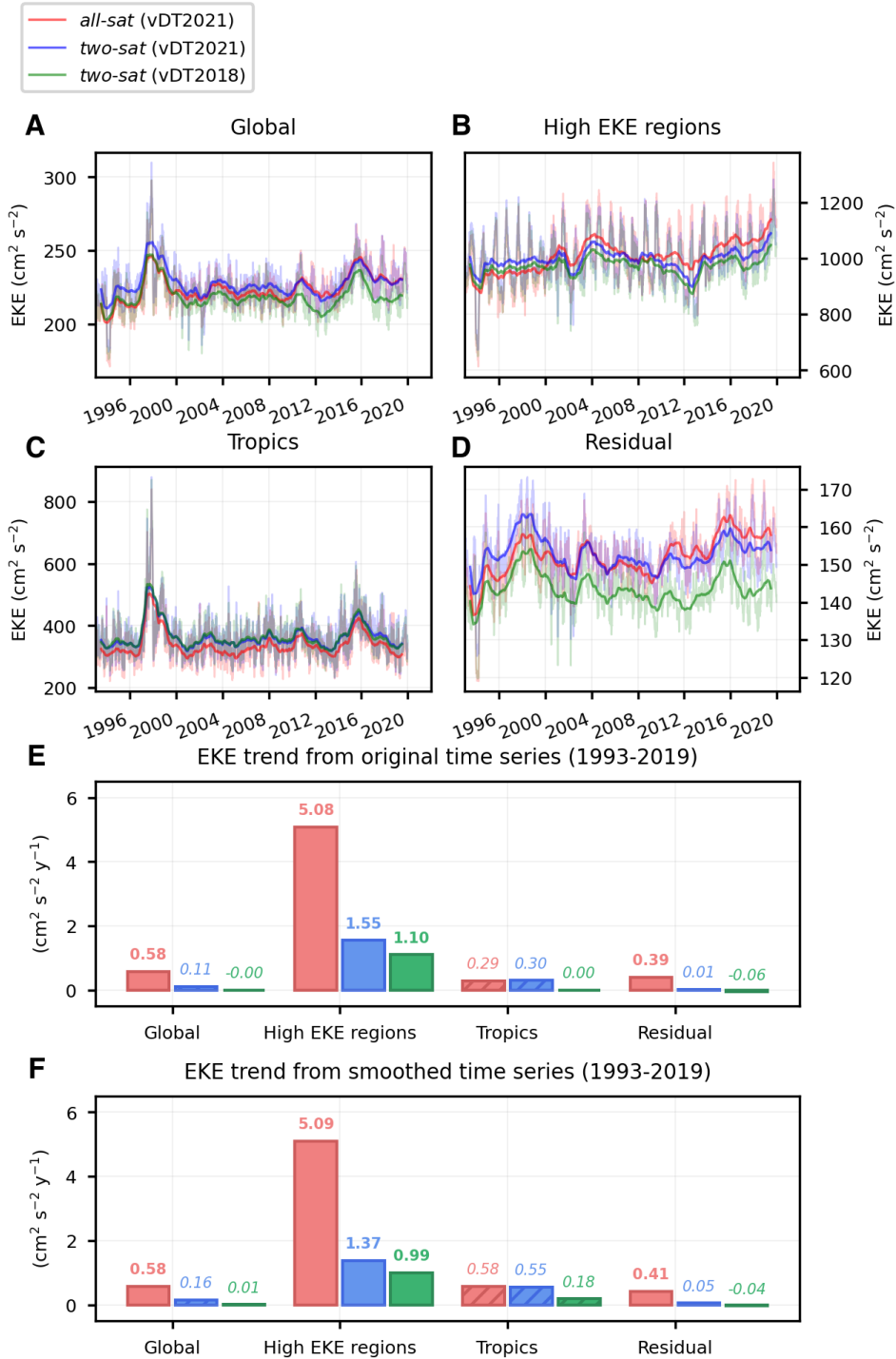

**Extended Data Fig. 3 | EKE time series and trends for three different altimetric products.**
EKE time series and their trends for the period between 1 January 1993 and 31 December 2019
(maximum available complete year for vDT2018) for three different altimetric products: *all-sat*
(vDT2021), *two-sat* (vDT2021), *two-sat* (vDT2018). Area-weighted mean EKE time series
computed over (**A**) the global ocean, (**B**) the high EKE regions, (**C**) the tropical band, and (**D**)
the global ocean excluding the high EKE regions and the tropical band (called residual), for the
*all-sat* vDT2021 (red line), *two-sat* vDT2021 (blue line), *two-sat* vDT2018 (green line) altimetric products. Thinner lines represent the original data, while thicker lines show the smoothed 365-day running average. (**E**) Trends of the original area-weighted mean EKE time series shown in (A-D). (**F**) Trends of the smoothed area-weighted mean EKE time series shown in (A-D). In (E) and (F) significant trends ($p < 0.05$) are written in bold text, while non-significant trends are written in italics.

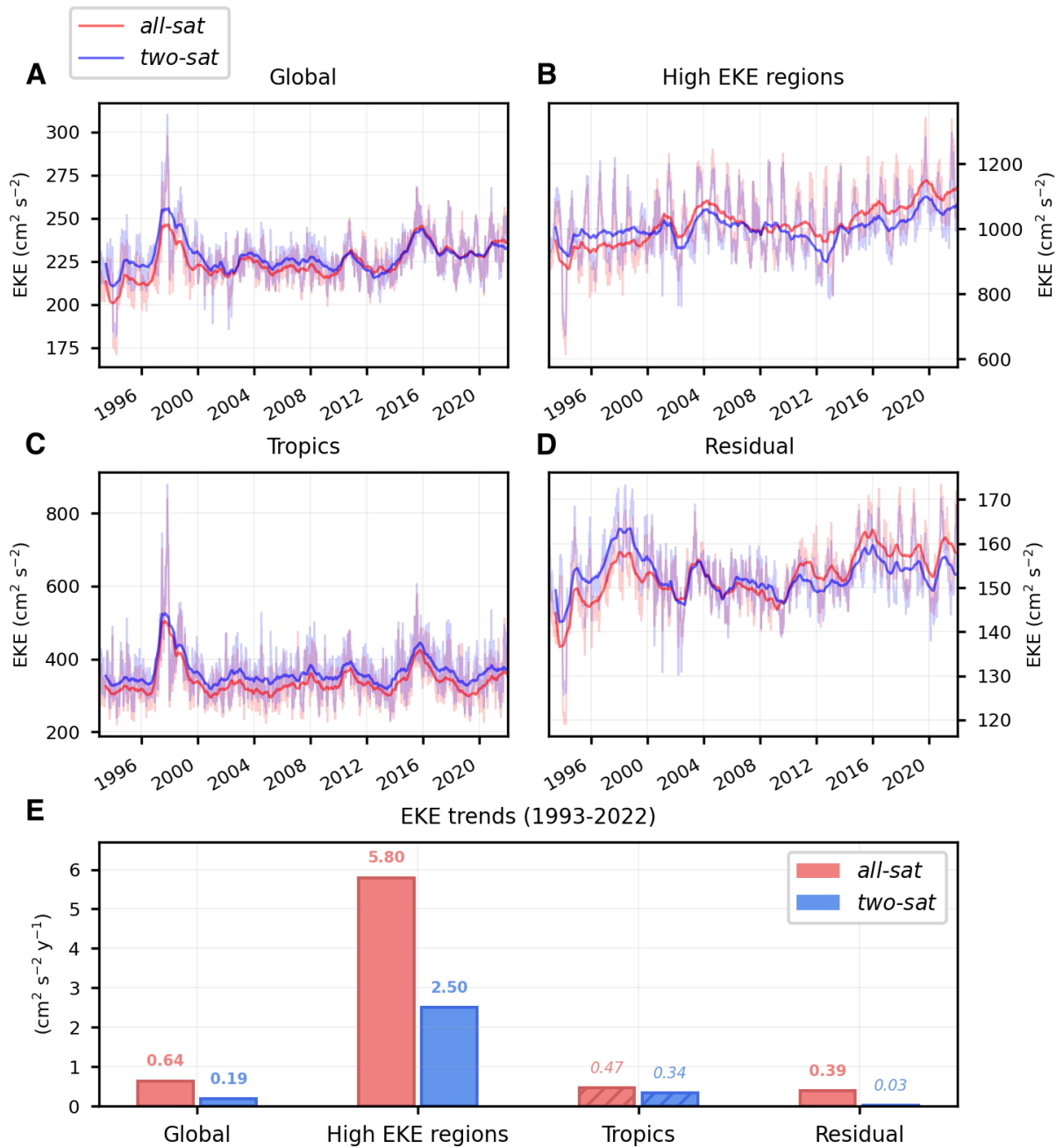

**Extended Data Fig. 4 | EKE time series and trends.** Area-weighted mean EKE time series computed over (**A**) the global ocean, (**B**) the high EKE regions, (**C**) the tropical band, and (**D**) the global ocean excluding the high EKE regions and the tropical band (called residual), for the *all-sat* (red line) and *two-sat* (blue line) altimetric products. Thinner lines represent the original data, while thicker lines show the yearly-rolling mean (i.e. 365-day-window moving average). (**E**) Trends of the original area-weighted mean EKE time series shown in (A-D). Significant trends (p < 0.05) are written in bold text, while non-significant trends are written in italics.

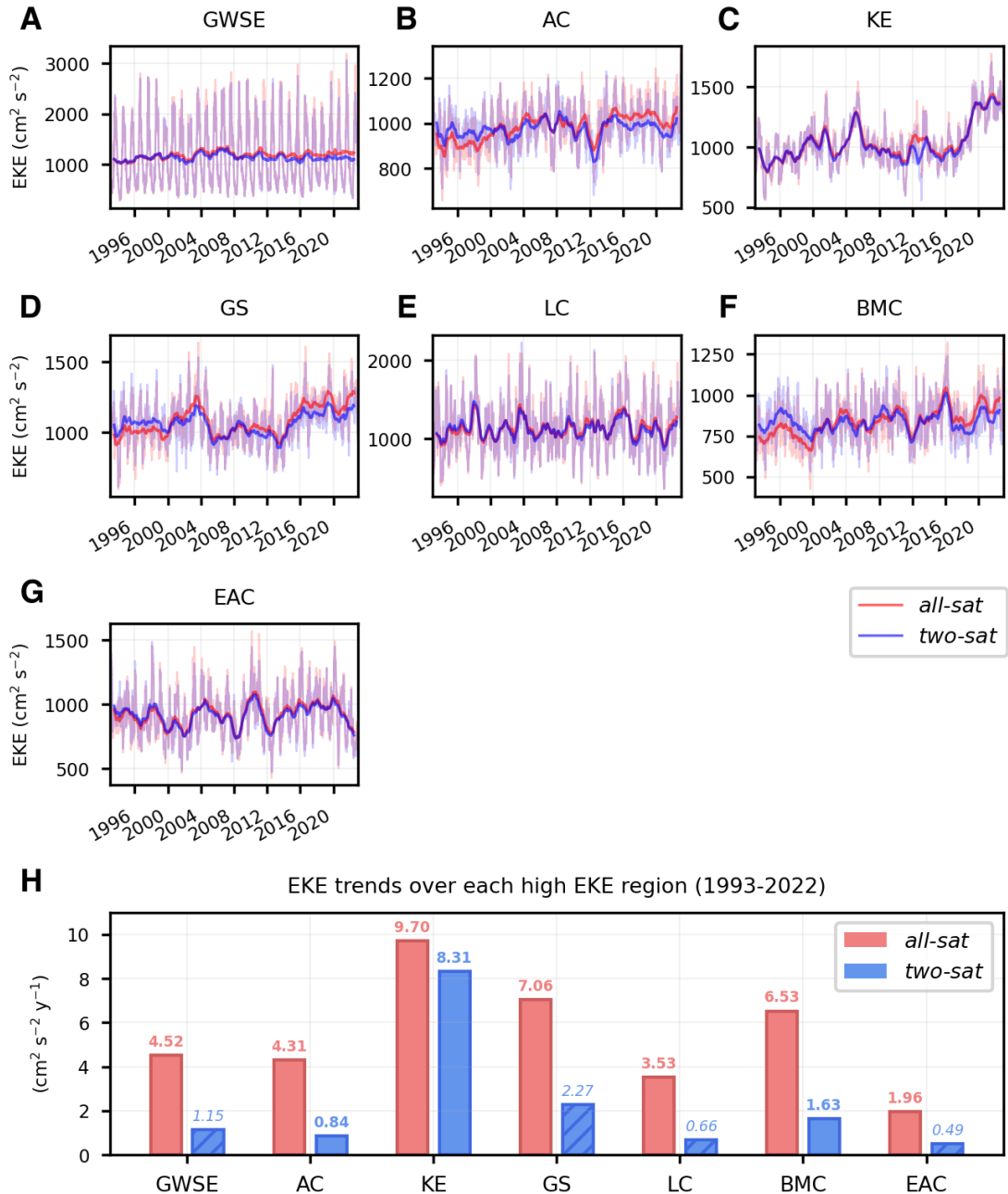

**Extended Data Fig. 5 | EKE time series and trends over each high EKE region.** Comparison of the area-weighted mean EKE time series computed over each high EKE region from 1993 to 2022: (**A**) Great Whirl and Socotra Eddy in East Africa (GWSE), (**B**) Agulhas Current (AC), (**C**) Kuroshio Extension (KE), (**D**) Gulf Stream (GS), (**E**) Loop Current (LC), (**F**) Brazil Malvinas Confluence region (BMC) and (**G**) East Australian Current (EAC). Thinner lines represent the original data, while thicker lines show the smoothed 365-day running average. (**H**) Trends of the

original area-weighted mean EKE time series shown in (A-G). Significant trends ($p < 0.05$) are written in bold text, while non-significant trends are written in italics.

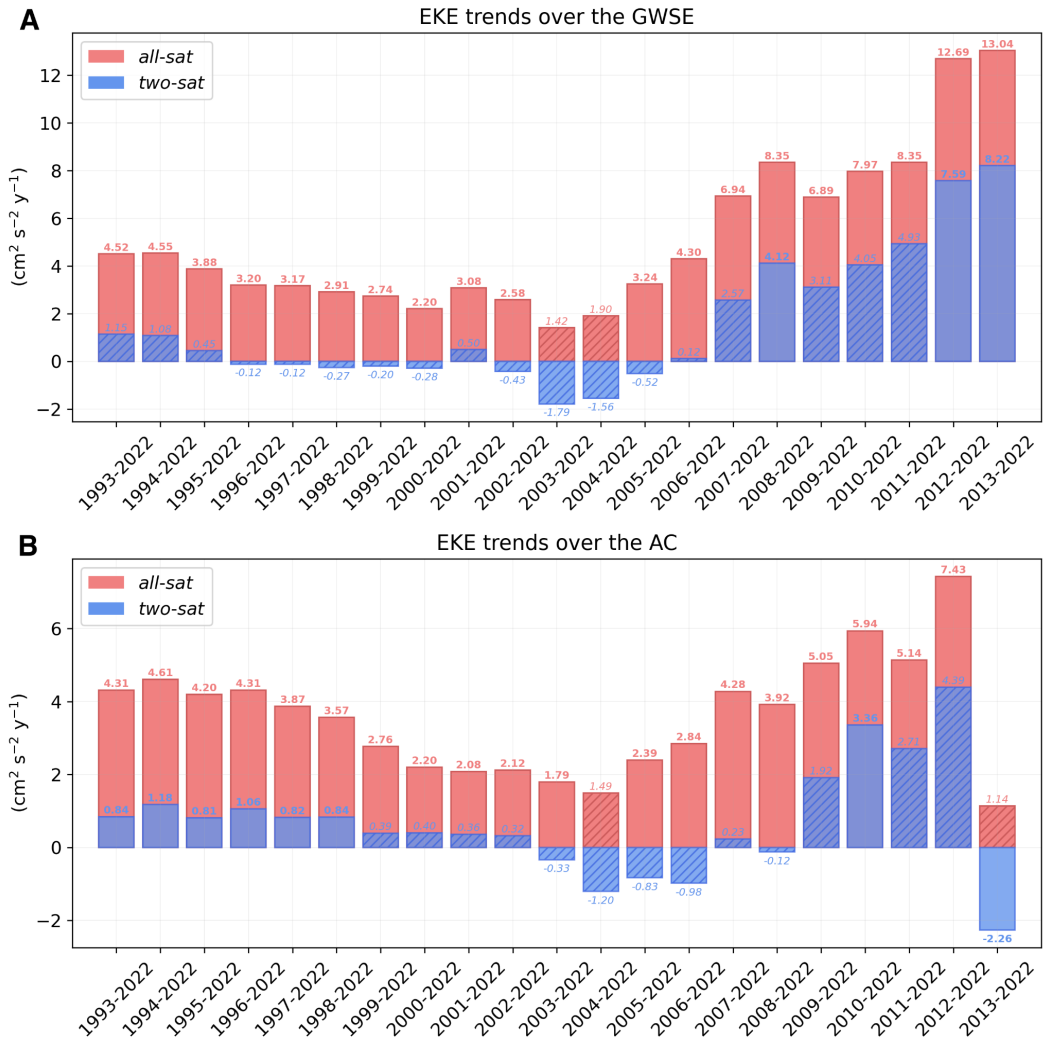

**Extended Data Fig. 6 (This figure continues on the next page) | Sensitivity test over each high EKE region.** Sensitivity test of the EKE trends computed over each high EKE region for different periods (Kuroshio Extension and Gulf Stream are shown in Fig. 4). Trends are computed from the original area-weighted mean EKE time series shown in Extended Data Fig. 5. Significant trends (p < 0.05) are written in bold text, while non-significant trends are written in italics. (**A**) Great Whirl and Socotra Eddy in East Africa (GWSE), (**B**) Agulhas Current (AC), (**C**) Loop Current (LC), (**D**) Brazil Malvinas Confluence region (BMC), and (**E**) East Australian Current (EAC).

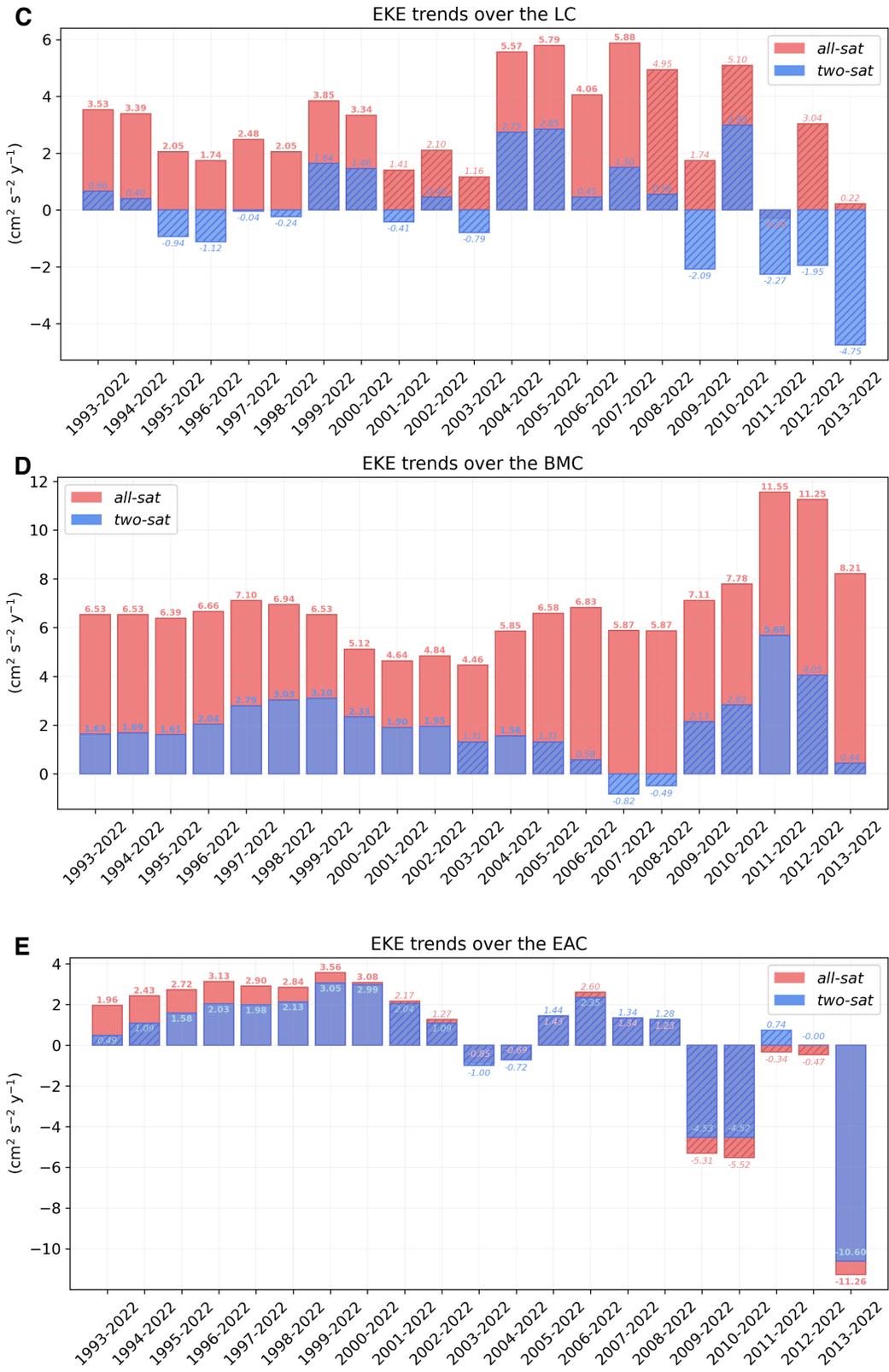

**Extended Data Fig. 6 (continuation)**

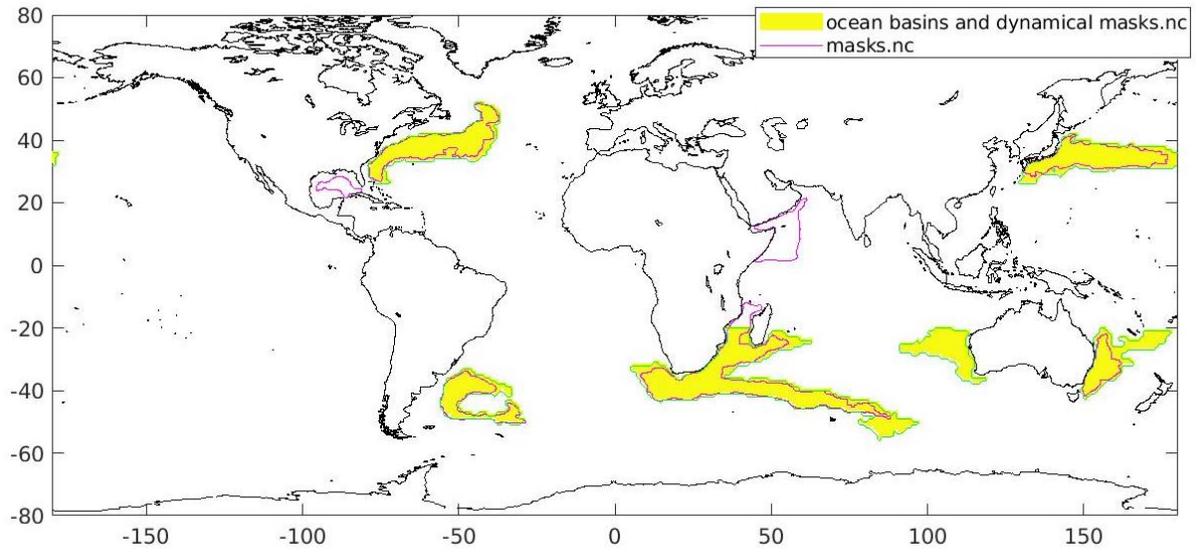

**Extended Data Fig. 7 | High EKE regions.** Masks of the high EKE regions used in our study (pink contours) and used by (*13*) (yellow regions).

| | Area ($km^2$) | Mean EKE ($cm^2\ s^{-2}$) | |
|---|---|---|---|
| | | *all-sat* | *two-sat* |
| Global | $3.28 \cdot 10^8$ | 224.83 | 227.09 |
| High EKE | $1.65 \cdot 10^7$ | 1018.12 | 1003.62 |
| Tropical | $4.39 \cdot 10^7$ | 330.51 | 354.68 |
| Residual | $2.77 \cdot 10^8$ | 152.73 | 152.46 |
| GWSE | $1.90 \cdot 10^6$ | 1179.16 | 1133.10 |
| AC | $5.92 \cdot 10^6$ | 982.23 | 974.27 |
| KE | $2.72 \cdot 10^6$ | 1043.04 | 1028.11 |
| GS | $2.80 \cdot 10^6$ | 1078.97 | 1061.14 |
| LC | $5.07 \cdot 10^5$ | 1133.63 | 1124.19 |
| BMC | $1.56 \cdot 10^6$ | 838.23 | 837.05 |
| EAC | $1.12 \cdot 10^6$ | 921.61 | 914.49 |

**Extended Data Table 1| Area and mean EKE over each region of study.** This table shows the total area of each study region (Area) and the area-weighted mean of the temporally averaged EKE computed over the period 1993-2022 from the *all-sat* and *two-sat* altimetric products (Mean EKE). The regions analyzed include the global ocean (Global), the high EKE regions (High EKE), the tropical band (Tropical), the global ocean excluding the high EKE regions and the tropical band (Residual), and each specific high EKE region: Gulf Stream (GS), Kuroshio Extension (KE), Agulhas Current (AC), Brazil Malvinas Confluence region (BMC), Loop Current (LC), Great Whirl and Socotra Eddy in East Africa (GWSE), and East Australian Current (EAC).